\documentclass{article}
\usepackage{spconf,graphicx,url}
\usepackage{amsmath,amsfonts,multirow,stmaryrd,dsfont}
\usepackage{algorithm,algorithmic}
\usepackage[justification=centering]{caption}
 
\graphicspath{{figures/}}
\title{Complex NMF under phase constraints based on signal modeling: application to audio source separation}

\name{Paul Magron \qquad Roland Badeau \qquad Bertrand David\thanks{This work is partly supported by the French National Research Agency (ANR) as a part of the the AIDA project (ANR-13-CORD-0001).}}

\address{LTCI, CNRS, T\'{e}l\'{e}com ParisTech, Universit\'{e} Paris-Saclay, 75013, Paris, France \\ \texttt{<firstname>.<lastname>@telecom-paristech.fr}}

\begin{document}
\ninept

\maketitle
\begin{abstract}
Nonnegative Matrix Factorization (NMF) is a powerful tool for decomposing mixtures of audio signals in the Time-Frequency (TF) domain. In the source separation framework, the phase recovery for each extracted component is necessary for synthesizing time-domain signals. The Complex NMF (CNMF) model aims to jointly estimate the spectrogram and the phase of the sources, but requires to constrain the phase in order to produce satisfactory sounding results. We propose to incorporate phase constraints based on signal models within the CNMF framework: a \textit{phase unwrapping} constraint that enforces a form of temporal coherence, and a constraint based on the \textit{repetition} of audio events, which models the phases of the sources within onset frames. We also provide an algorithm for estimating the model parameters. The experimental results highlight the interest of including such constraints in the CNMF framework for separating overlapping components in complex audio mixtures.
\end{abstract}
\begin{keywords}
Nonnegative matrix factorization, phase recovery, phase unwrapping, repeated audio events, source separation
\end{keywords}
\section{Introduction}
\label{sec:intro}

A variety of audio signal processing techniques acts in the TF domain, exploiting the particular structure of music signals. For instance, the family of techniques based on Nonnegative Matrix Factorization (NMF)~\cite{Lee1999} is often applied to spectrogram-like representations, and has proved to provide a successful and promising framework for audio source separation~\cite{Smaragdis2003}.

However, when it comes to resynthesizing time signals, obtaining the phase of the corresponding Short-Time Fourier Transform (STFT) is necessary, and is still an open issue~\cite{Gerkmann2015}. In order to produce perceptually satisfactory sounding signals, it is important to enforce \emph{consistency}, i.e. to obtain a complex-valued component that is close to the STFT of a time signal. In the source separation framework, a common practice consists in applying Wiener-like filtering~\cite{Fevotte2009}. However, this method does generally not lead to consistent components. Alternatively, a consistency-based approach is often used for phase recovery~\cite{Griffin1984}. That is, a complex-valued matrix is iteratively computed in order to maximize its consistency. A benchmark has been conducted to assess the potential of source separation methods with phase recovery in NMF~\cite{Magron2015}. It points out that consistency-based approaches provide poor results in terms of audio quality. Besides, Wiener filtering fails to provide good results when sources overlap in the TF domain. The High Resolution NMF (HRNMF) model~\cite{Badeau2014} has shown a great potential because it relies on a signal model, contrary to consistency-based approaches that exploit a property of the TF transform. The Complex NMF (CNMF) model~\cite{Kameoka2009}, which consists in factorizing a magnitude spectrogram while reconstructing a phase field for each source, has also shown promising results, but requires that the phase be constrained to produce satisfactory results.


Alternatively, phase reconstruction from spectrograms can be performed using phase models based on signal analysis. For instance, the widely used model of mixtures of sinusoids~\cite{McAuley1986} can lead to explicit constraints for phase reconstruction that exploit the relationships between adjacent TF bins. This approach has been used in the phase vocoder algorithm~\cite{Flanagan1966}, integrated into a CNMF framework~\cite{Bronson2014} and applied to speech signal reconstruction~\cite{Krawczyk2014}. In~\cite{Magron2015a}, we proposed to generalize this approach and have provided a phase unwrapping algorithm that has been applied to an audio signal restoration task. Thus, modeling the phase of the components in the TF domain by means of a signal model appears to be a promising approach. In addition to this model, one can also exploit the repetition of audio events to obtain a phase constraint in the TF domain. A phase reconstruction technique exploiting this property has been proposed in~\cite{Magron2015c} for estimating the phases of the components from the mixture within onset frames. This method has been applied to audio source separation and has shown some potential, but the spectrograms of the isolated components were assumed to be known.

In this paper, we propose to define a mixture model that incorporates the aforementioned phase constraints for jointly estimating both the magnitude and the phase of the components. The mixture is decomposed into several sources, whose STFTs are structured according to the CNMF model. In addition, we include the following constraints: a \textit{phase unwrapping} model that enforces the continuity of the partials over time frames, and a model based on the \textit{repetition} of audio events for estimating the phases of the components within onset frames. We incorporate these constraints into the CNMF model by means of penalty functions in the complete cost function, and we minimize it in order to derive an estimation algorithm. This method is tested experimentally on various data and applied to an audio source separation task.

This paper is organized as follows. Section \ref{sec:background} presents the necessary background on NMF, CNMF and phase-constrained CNMF. Section \ref{sec:model} introduces our model and the derived estimation algorithm. Section \ref{sec:exp} describes several experiments that highlight the potential of this technique. Finally, section \ref{sec:conclu} draws some concluding remarks and prospects future directions.

\section{Background}
\label{sec:background}

\subsection{Nonnegative Matrix Factorization}

The NMF problem is expressed as follows: given a matrix $V$ of dimensions $F \times T$ with nonnegative entries, find a factorization $V \approx \hat{V} = WH$ where $W$ and $H$ are nonnegative matrices of dimensions $F \times K$ and $K \times T$. In order to reduce the dimension of the data, $K$ is chosen such that $K(F+T)\ll FT $. In audio source separation, $V$ is generally the magnitude or the power spectrogram of a TF representation $X$ of a mixture signal (most of the time an STFT). One can interpret $W$ as a dictionary of spectral templates and $H$ as a matrix of temporal activations.


This factorization is generally obtained by minimizing a cost function $D(V,\hat{V})$. Popular choices for $D$ are the Euclidean distance and the Kullback-Leibler~\cite{Lee1999} or the Itakura-Saito~\cite{Fevotte2009} divergence.

Finally, the phase of the complex components must be recovered in order to resynthesize time-domain signals. A common practice consists in applying Wiener filtering~\cite{Fevotte2009}, which corresponds to a soft-masking of the complex-valued STFT of the original mixture. Thus, the phase of each source is equal to the phase of the mixture. However, this property is no longer valid when sources overlap in the TF domain, which motivates alternative methods for reconstructing the phases of the components~\cite{Griffin1984}.

\subsection{Complex NMF}

CNMF \cite{Kameoka2009} consists in factorizing a magnitude spectrogram while reconstructing a phase field for each source from the STFT $X$ of the mixture. The model is, $\forall (f,t)\in \llbracket 0;F-1 \rrbracket \times \llbracket 0;T-1 \rrbracket $\footnote{The notation $\llbracket \text{ } \rrbracket$ denotes an integer interval.}:

\begin{equation}
\hat{X}(f,t) = \sum_{k=1}^{K} W(f,k) H(k,t) e^{i\phi_k(f,t)}.
\end{equation}

The model parameters are estimated by minimizing the squared Euclidean distance between the model $\hat{X}$ and the data $X$:

\begin{equation}
D(X,\hat{X}) = ||X-\hat{X}||^2 = \sum_{f,t} |X(f,t) - \hat{X}(f,t) | ^2.
\label{eq:cost_euclidean}
\end{equation}

A penalty term is generally added to this cost function in order to promote sparse activations~\cite{Hoyer2004}. The CNMF model~\cite{Kameoka2009} uses the following sparsity penalty term:

\begin{equation}
\mathcal{C}_s(H) = 2 \sum_{k,t}H(k,t)^p,
\label{eq:cost_sparse}
\end{equation}
where $p$ is a sparsity parameter chosen between $0$ and $2$. More details on the estimation procedure can be found in~\cite{Sawada2011}.

\subsection{Phase-constrained Complex NMF}

Although promising, the Complex NMF model has been shown to provide poor results in terms of audio source separation quality because the phase is left unconstrained~\cite{Magron2015}. In~\cite{LeRoux2009a}, the authors proposed to enforce the \textit{consistency} of the estimated components, but this property was shown to be uncorrelated to the sounding quality.

An alternative approach consists in using phase constraints based on signal modeling. In~\cite{Bronson2014}, the authors proposed a temporal phase evolution constraint in the TF domain, based on the explicit calculation of the phase of mixtures of sinusoids. Although promising, this approach is limited to harmonic and stationary signals, and requires prior knowledge on fundamental frequencies and numbers of partials. Thus, it is not suitable for blind source separation.

Finally, time invariant parameters can be exploited to constrain the phase within a complex NMF framework. In~\cite{Kirchhoff2014}, the authors proposed to use the phase offset between partials in order to reduce the dimensionality of the data. However, the corresponding estimation algorithm is not able do deal with a mixture of different instrumental sources.

\section{Proposed model}
\label{sec:model}

Drawing on previous work~\cite{Magron2015a,Magron2015c}, we propose to incorporate the following constraints into a CNMF framework:

\begin{itemize}
\item A \textit{phase unwrapping} constraint, that enforces the continuity of the partials composing the spectra of the sources over time frames;
\item A phase model based on the \textit{repetition} of audio events in order to estimate the phases of the sources within onset frames.
\end{itemize}

In this paper, we assume that the onset frame indexes are known. However,
we could implement, for instance, the method described in~\cite{Paulus2005} for estimating onset frames from the activation matrix $H$.

\subsection{Phase unwrapping}

Let us consider a source indexed by $k \in \llbracket 1 ; K \rrbracket$ which is modeled as a sum of sinusoids. $\nu_k(f)$ denotes the normalized frequency in channel $f \in \llbracket 0 ; F-1 \rrbracket$. We define the onset domain for each source: 

\begin{equation}
\Omega_k = \{  t \in \llbracket 0 ; T-1 \rrbracket, t \text{ is an onset frame index for source } k  \}.
\label{eq:onset_domain}
\end{equation}

It can be shown~\cite{Magron2015a} that the phase $\phi_k$ of the component $X_k$ follows the unwrapping equation: $\forall f \in \llbracket 0 ; F-1 \rrbracket $ and $\forall t \notin \Omega_k$,

\begin{equation}
\phi_k(f,t) = \phi_k(f,t-1) + 2 \pi S \nu_k(f),
\label{eq:phase_unwrapping}
\end{equation}
where $S$ is the hop size of the STFT. From~\eqref{eq:phase_unwrapping} we define the unwrapping cost function:

\begin{equation}
\mathcal{C}_u(\phi) = \sum_{f,k} \sum_{t \notin \Omega_k }  |X(f,t)|^2 u_k(f,t),
\label{eq:cost_unwrapping}
\end{equation}
with  $ u_k(f,t) =| e^{i\phi_k(f,t)}e^{-i\phi_k(f,t-1)} - e^{2i \pi S \nu_k(f) }   |^2$. The relative importance of the constraint is weighted by the terms $|X(f,t)|^2$.

In~\cite{Bronson2014} the frequencies $\nu_k(f)$ were assumed to be known. We propose here to estimate them with a Quadratic Interpolated FFT (QIFFT) performed on the spectra $W_k$. The frequency range is then decomposed into \textit{regions of influence}~\cite{Magron2015a} to ensure that the phase in a given channel is unwrapped with the appropriate frequency.

\subsection{Phase constraint based on a model of repeated audio events}

Another constraint can be added in order to reconstruct the phase of the sources within onset frames. Indeed, reconstructing a good quality onset signal is crucial since it has a significant impact on the sound quality and because it initializes the recursive equation~\eqref{eq:phase_unwrapping}.

We propose to use a phase model based on the repetition of audio events~\cite{Magron2015c}. The phase of a source within an onset frame is modeled as the sum of a \textit{reference} phase and an \textit{offset} which is linearly dependent on the frequency: $\forall  k \in \llbracket 1 ; K \rrbracket$, $f \in \llbracket 0 ; F-1 \rrbracket$ and $t \in \Omega_k$,

\begin{equation}
\phi_k(f,t) = \psi_k(f) + \lambda_k(t) f.
\label{eq:phase_rep}
\end{equation}
This model leads to the following cost function:

\begin{equation}
\mathcal{C}_r(\phi,\psi,\lambda) = \sum_{f,k} \sum_{t \in \Omega_k }  |X(f,t)|^2 r_k(f,t),
\label{eq:cost_rep}
\end{equation}
with $ r_k(f,t) = | e^{i\phi_k(f,t)} - e^{i\psi_k(f)} e^{i \lambda_k(t) f }   |^2$.

\subsection{CNMF under phase constraints}

The aforementioned constraints are incorporated into the CNMF framework by adding the penalty terms~\eqref{eq:cost_sparse}, \eqref{eq:cost_unwrapping} and \eqref{eq:cost_rep} to the cost function~\eqref{eq:cost_euclidean}, which leads to the following objective function:

\vspace{-1em}
\begin{equation}
\mathcal{C}(\theta) =  D(X,\hat{X}) + \sigma_u \mathcal{C}_u(\phi) + \sigma_r \mathcal{C}_r(\phi,\psi,\lambda) +   \sigma_s \mathcal{C}_s(H),
\label{eq:cnmf_phase_cost}
\end{equation}
where $\theta = \{ W,H,\phi,\psi,\lambda \}$ and the parameters $\sigma_u$, $\sigma_r$ and $\sigma_s$ are prior weights which promote the constraints.

The cost function is minimized by means of a coordinate descent method. We calculate the partial derivative of $\mathcal{C}$ with respect to the parameters $\{ W,H,\phi,\psi\}$ and we seek them such that this derivative is zero. The parameters $\lambda$ are estimated by means of an adaptation of the ESPRIT algorithm~\cite{Hua2004}. Besides, we enforce the nonnegativity of $W$ and $H$ by a projection onto nonnegative orthants, and we add some normalization. Thus, the convergence of the algorithm is not theoretically guaranteed, however it was observed in our experiments. Algorithm~\ref{al:cnmf_ph} describes the estimation procedure. We use the following notations: $\forall  k \in \llbracket 1 ; K \rrbracket$,

\vspace{-1em}
\begin{align*}
\mu_k \in \mathbb{C}^{F \times 1} &\text{, } \mu_k(f) = e^{2i \pi S \nu_k(f)}, \\
\Lambda_k \in \mathbb{C}^{F \times T} &\text{, } \Lambda_k(f,t) = \mathds{1}_k(t) e^{i f \lambda_k(t)}, \\
\Psi_k \in \mathbb{C}^{F \times 1} &\text{, } \Psi_k(f) = e^{ i  \psi_k(f)}, \\
\Phi_k \in \mathbb{C}^{F \times T} &\text{, } \Phi_k(f,t) = e^{i \phi_k(f,t)},
\end{align*}
and let $\alpha \in \mathbb{C}^{F \times T} \text{, } \alpha(f,t)=1$. The indicator function of $\Omega_k$ is:

\begin{equation*}
\mathds{1}_k \in \mathbb{C}^{1 \times T} \text{, }
\mathds{1}_k(t)=
\left\lbrace
\begin{array}{cc}
1 & \text{ if } t \in \Omega_k \\
0 & \text{ else.}
\end{array}\right.
\label{eq:ind_omega}
\end{equation*}
and the indicator function of the complementary set is $ \widetilde{\mathds{1}}_k = 1 - \mathds{1}_k$. $\Re$ denotes the real part, $\mathtt{vand}(z)$ (resp. $\mathtt{diag_v}(v)$) denotes the Vandermonde matrix\footnote{If $z \in \mathbb{C}^{1 \times T}$, then $M=\mathtt{vand}(z) \in \mathbb{C}^{F \times T}$. This Vandermonde matrix is definied as follows: $\forall (f,t)\in \llbracket 0;F-1 \rrbracket \times \llbracket 0;T-1 \rrbracket $, $M(f,t) = z(t)^f$.} (resp. the diagonal matrix) made up with the entries of vector $v$ and $\mathtt{diag_m}(M)$ denotes the column vector made up with the diagonal entries of matrix $M$. $z_\downarrow$ (resp. $z_\uparrow$) denotes the vector or matrix obtained by removing the last (resp. the first) entry from vector or matrix $z$. $M_{0 \rightarrow}$ denotes the concatenation of a column vector whose entries are zeros and the matrix obtained by removing the last column from $M$. Finally, $.^T$ (resp. $\overline{.}$ and $.^H$) denotes the transpose (resp. the conjugate and the Hermitian transpose), and $\odot$ (resp. $\frac{.}{.}$ and $.^{\odot}$) denotes the element-wise matrix multiplication (resp. division and power).

\begin{algorithm}[t!]
\caption{CNMF under phase constraints}
\label{al:cnmf_ph}
\begin{algorithmic}

\STATE \textbf{Inputs}: \\
$X$, $K$, $\sigma_r$, $\sigma_u$, $\sigma_s$

\STATE \textbf{Initialization $\forall k \in \llbracket 1;K \rrbracket $}: \\
$W_k$, $H_k$, $\Phi_k$, $\Lambda_k$, $\Psi_k$, $\mathds{1}_k$, $\overline{\mathds{1}}_k$, \\
$ \hat{X}_k = (W_k H_k) \odot \Phi_k$, $ B_k = X - \sum_{l \neq k} \hat{X}^l $.

\WHILE{stopping criteria not met}

\FOR{$k=1 \text{ to } K$}

\item \textbf{Compute $\mu_k$} \\

QIFFT on $W_k$~\cite{Magron2015a}.

\item \textbf{Compute $\Psi_k$}

$\Psi_k = \dfrac{\mathtt{diag_m}( (\Phi_k  \odot X^{\odot 2}) (\Lambda_k)^H )}{|\mathtt{diag_m}( (\Phi_k  \odot X^{\odot 2}) (\Lambda_k)^H )|}.$

\item \textbf{Compute $\Lambda_k$}

$\Lambda_k = \mathtt{vand} \left( \dfrac{(\overline{\Psi}_{k,\downarrow} \odot \Psi_{k,\uparrow} )^H ( |X_{k,\downarrow}| \odot |X_{k,\uparrow}| \odot \overline{\Phi}_{k,\downarrow} \odot \Phi_{k,\uparrow})}{|(\overline{\Psi}_{k,\downarrow} \odot \Psi_{k,\uparrow} )^H ( |X_{k,\downarrow}| \odot |X_{k,\uparrow}| \odot \overline{\Phi}_{k,\downarrow} \odot \Phi_{k,\uparrow})|} \right).$ \\

$\Lambda_k = \Lambda_k \mathtt{diag_v}(\mathds{1}_k).$

\item \textbf{Compute $\Phi_k$}

$\rho_k = \sigma_r (\Psi_k \mathds{1}_k) \odot \Lambda_k +  \sigma_u (\mu_k \widetilde{\mathds{1}}_k) \odot \Phi_{k,0 \rightarrow}.$

$\Phi_k =  \dfrac{B_k \odot (W_k H_k) + (W_k H_k)^{\odot 2} \odot \rho_k}{|B_k \odot (W_k H_k) + (W_k H_k)^{\odot 2} \odot \rho_k|}.$

\item \textbf{Update $\hat{X}_k$}

$ \hat{X}_k = (W_k H_k) \odot \Phi_k$, \\
$ B_k = X - \sum_{l \neq k} \hat{X}^l $,\\
$\beta_k = \Re( B_k \odot \overline{\Phi}_k )$.

\item \textbf{Compute $W$}

$W_k = \dfrac{\beta_k (H_k)^T}{\alpha ((H_k)^{\odot 2})^T}$.

\item \textbf{Project $W$ onto nonnegative orthant}

\item \textbf{Compute $H$}

$H_k = \dfrac{(W_k)^T \beta_k}{ p \sigma_s (H_k)^{\odot p-2} + ((W_k)^{\odot 2})^T \alpha}$.

\item \textbf{Project $H$ onto nonnegative orthant}

\item \textbf{Normalize $W$ and $H$}

\item \textbf{Update $\hat{X}_k$}

$ \hat{X}_k = (W_k H_k) \odot \Phi_k$, \\

$ B_k = X - \sum_{l \neq k} \hat{X}^l $.

\ENDFOR
\ENDWHILE
\STATE \textbf{Outputs}: \\
$\hat{X}_k$, $W_k$, $H_k$, $\Phi_k$, $\Lambda_k$, $\Psi_k$, $\mu_k$.

\end{algorithmic}
\end{algorithm}

\section{Experimental evaluation}
\label{sec:exp}

\subsection{Protocol and datasets}

We perform audio source separation on several datasets. First, we synthesize $30$ mixtures of two harmonic signals ($K=2$) which consist of damped sinusoids whose amplitude, origin phase, frequency and damping coefficients are randomly-defined. For the tests on realistic data, we consider $30$ mixtures of two piano notes ($K=2$) selected randomly from the MAPS database~\cite{Emiya2010a}. In both datasets, sources overlap in the TF domain. Each source is activated alone successively, and then both are played simultaneously.

The data is sampled at $F_s = 11025$ Hz. The STFT is calculated with a $512$ sample-long modified Hann window (as defined in~\cite{Griffin1984}) with $75 \%$ overlap. In order to measure the quality of the source separation, we use the \textsc{BSS Eval} toolbox~\cite{Vincent2006} which computes various energy ratios: the Signal to Distortion, Interference and Artifact Ratios (SDR, SIR and SAR).

We test various methods: \textbf{NMF-W} consists of 30 iterations of NMF with Kullback-Leibler divergence (KLNMF) multiplicative update rules combined with Wiener filtering~\cite{Fevotte2009}; \textbf{CNMF} consists of 10 iterations of the sparse CNMF algorithm without phase constraint~\cite{Kameoka2009}; Finally, \textbf{CNMF-$\phi$} consists of 10 iterations of Algorithm~\ref{al:cnmf_ph}. For both CNMF methods, $W$, $H$ and $\phi$ are initialized with $30$ iterations of \textbf{NMF-W} and the other parameters are initialized randomly.

\subsection{Influence of the weight parameters}

\begin{figure}[t]
\includegraphics[scale=0.33]{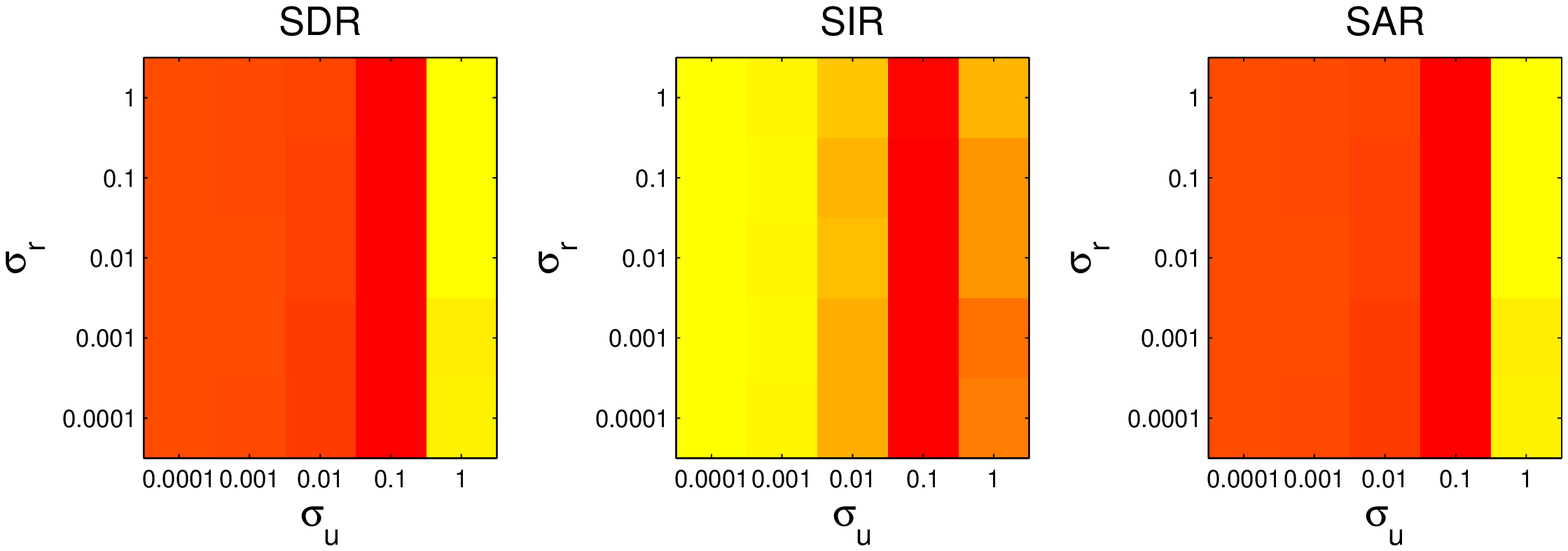}
\includegraphics[scale=0.33]{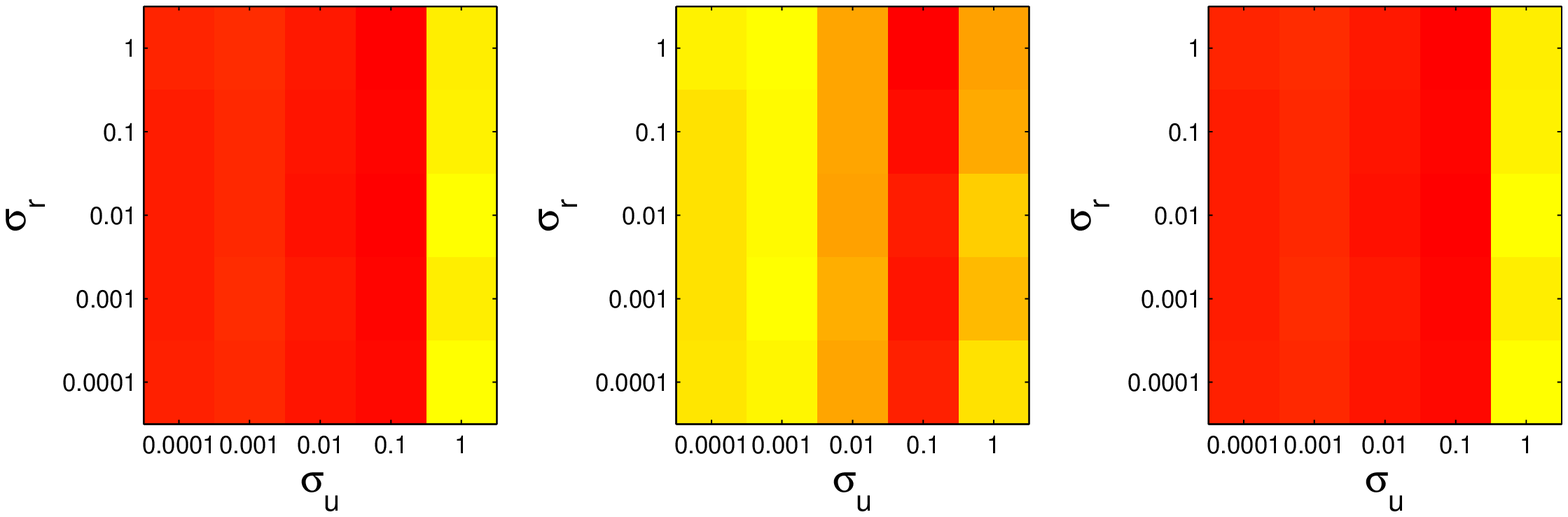}
\vspace{-2em}
\caption{Influence of the parameters $\sigma_u$ and $\sigma_r$ on the source separation quality. The dark red color corresponds to the greatest values of the SDR, SIR and SAR. Results on synthetic sinusoids (top) and piano notes (bottom).}
\label{fig:inf_sigma}
\end{figure}

\begin{figure}[t]
\centering
\includegraphics[scale=0.4]{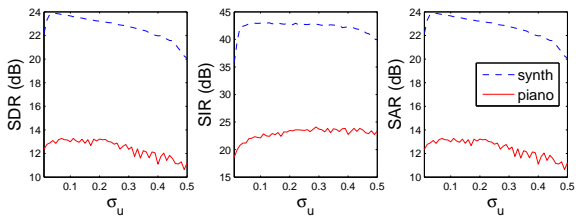}
\vspace{-2em}
\caption{Influence of the parameter $\sigma_u$ on the separation quality for the two datasets, with $\sigma_r = 0.2$.}
\vspace{-1em}
\label{fig:inf_sigma_u}
\end{figure}

The first experiment analyzes the influence of the unwrapping and repetition parameters $\sigma_u$ and $\sigma_r$ on the performance of the source separation based on the CNMF under phase constraints method. The sparsity parameters are set at $p=1$ and $\sigma_s = ||X||^2 K^{-(1-p/2)} 10^{-5} $, which are commonly used values in the literature~\cite{Kameoka2009}. We run the \textbf{CNMF-$\phi$} procedure with various values of $\sigma_r$ and $\sigma_u$. We then compute the SDR, SIR and SAR and average the results over each dataset. Results are presented in figure~\ref{fig:inf_sigma}. We remark that $\sigma_r$ does not have a great influence on the quality of the separation. However, the quality significantly drops for values of $\sigma_r$ and $\sigma_u$ higher than $1$. We thus set $\sigma_r = 0.2$ and investigate more precisely the influence of $\sigma_u$, as presented in figure~\ref{fig:inf_sigma_u}. We remark that small values of $\sigma_u$ ($\approx 0.05$) lead to the best results for synthetic data. However, values in the neighborhood of $0.2$ seem to lead to the most stable results for real piano notes. Thus, we will use the values $(\sigma_u,\sigma_r) = (0.2 , 0.2)$ in the next tests in order to guarantee as much robustness as possible in our experimental results.

\subsection{Source separation}
\begin{figure}[t]
\begin{minipage}[b]{.48\linewidth}
  \centering
  \centerline{\includegraphics[scale=0.3]{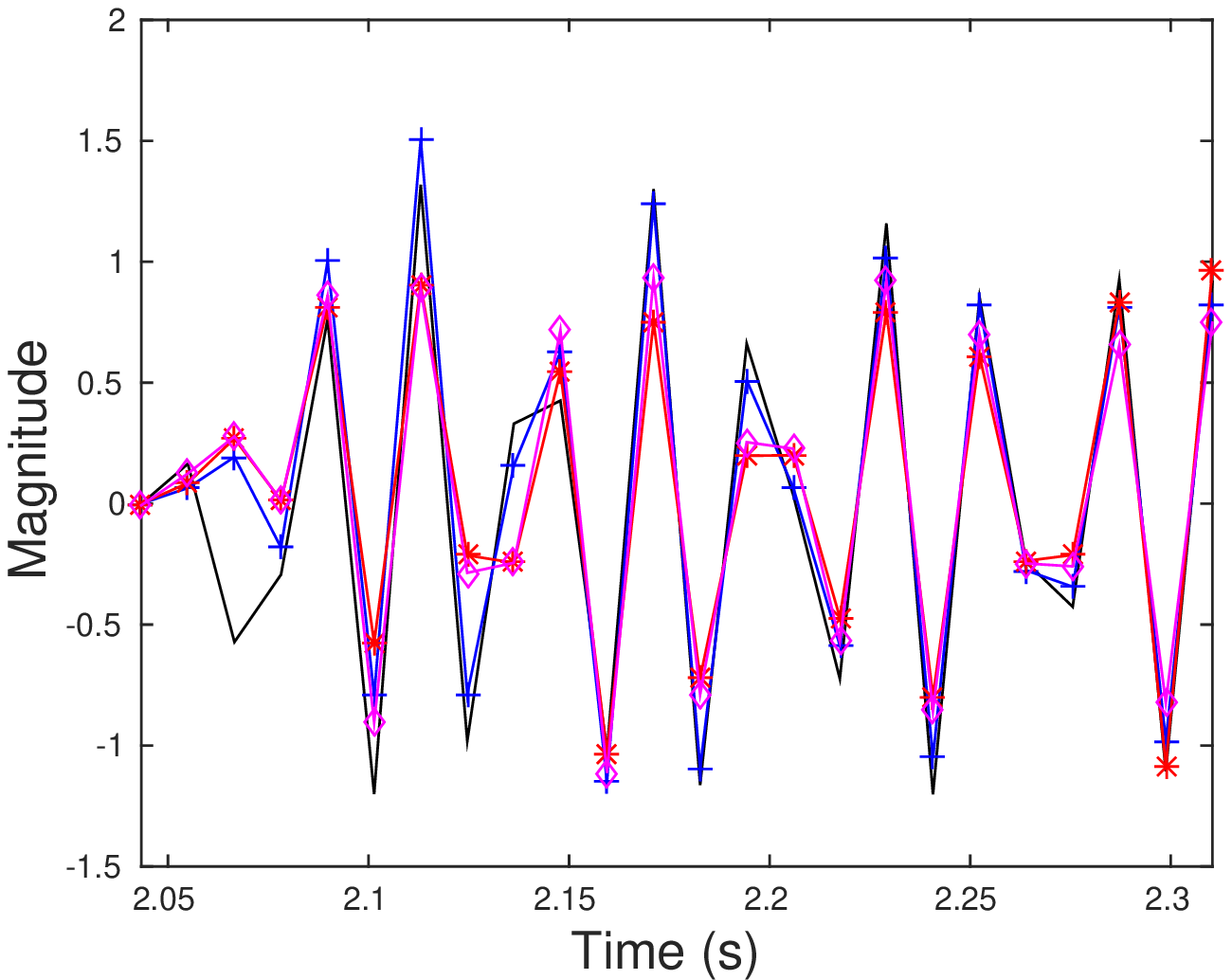}}
  \centerline{(a) Real part}\medskip
\end{minipage}
\hfill
\begin{minipage}[b]{0.48\linewidth}
  \centering
  \centerline{\includegraphics[scale=0.3]{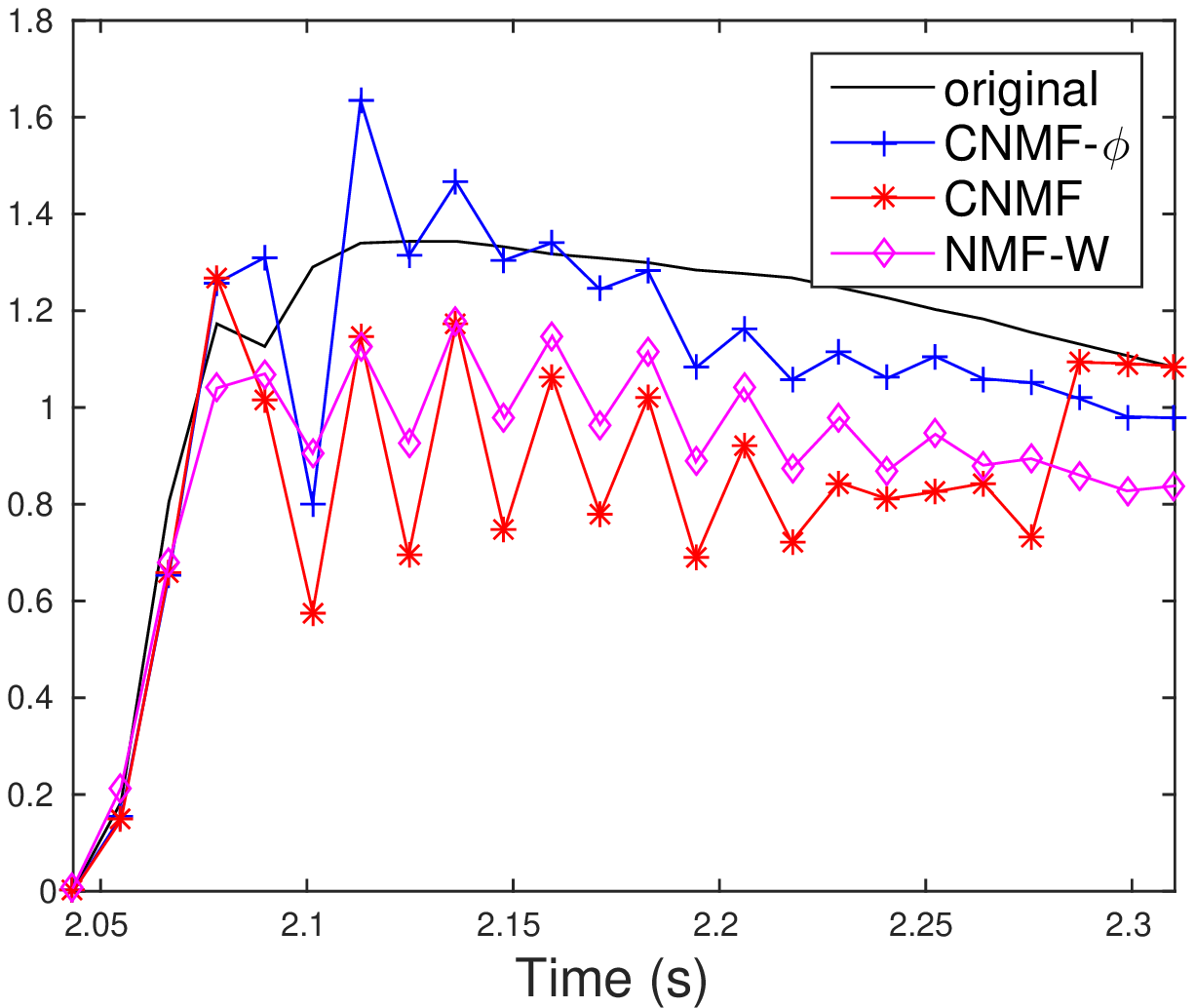}}
  \centerline{(b) Magnitude}\medskip
\end{minipage}
\vspace{-1em}
\caption{Reconstruction of a B2 piano note partial from a mixture made up of two piano notes (E2 and B2) with different methods.}
\label{fig:cnmf_E2B2}
\end{figure}

As a first example, we consider a mixture of two piano notes that overlap in the TF domain (E2 and B2). Each note is played alone successively, and then both are played simultaneously. We perform source separation with the methods described in the protocol. We then look at the reconstructed component corresponding to the note B2 in the $58$-th frequency channel between $2$ s and $2.3$ s, i.e. when overlap between partials of the two piano notes occurs. Results presented in figure~\ref{fig:cnmf_E2B2} show that our method accurately reconstructs the piano partial in this frequency channel. Both the real part and the magnitude of the components are well reconstructed with our technique, while the other methods lead to unsatisfactory results.

\begin{table}[t]
\center
\begin{tabular}{c|c||c|c|c}
Data & Method & SDR & SIR & SAR \\
\hline
\multirow{2}{*}{Synthetic sinsuoids}
& NMF-W  & $12.1$  & $17.5$ & $14.1$ \\
& CNMF   & $12.0$  & $14.6$ & $\mathbf{16.1}$  \\
& CNMF-$\phi$   & $\mathbf{14.0}$  & $\mathbf{20.7}$ & $15.4$  \\
 \hline
\multirow{2}{*}{Piano notes} 
& NMF-W  & $12.9$  & $23.3$ & $14.5$ \\
& CNMF   & $13.5$  & $20.0$ & $\mathbf{14.8}$  \\
& CNMF-$\phi$   & $\mathbf{14.0}$  & $\mathbf{24.0}$ & $14.6$  \\
 \hline
\end{tabular}
\vspace{-0.5em}
\caption{Average source separation performance (SDR, SIR and SAR in dB) for various methods and data.}
\label{tab:cnmf_ph_perf}
\vspace{-1em}
\end{table}

Finally, we perform source separation on the two datasets and average the source separation scores over the mixtures. We present the results in Table~\ref{tab:cnmf_ph_perf}. Our method outperforms the Wiener filtering technique in terms of interference rejection. A slight increase in SDR is also observed on both datasets, while the CNMF technique leads to the best performance in terms of artifact rejection. An informal perceptive evaluation of the source separation quality suggests that the performance measurement employed in these tests may not be able to capture some properties of the separated signals. In particular, the beating phenomenon cannot be suppressed when the phase is retrieved with Wiener filtering, which significantly impacts the sounding quality of the reconstructed signal, while our technique dramatically attenuates this phenomenon. In order to assess the potential of this method, we provide on our webpage~\cite{Magron} some audio examples, as well as the MATLAB code related to this work.

\section{Conclusion}
\label{sec:conclu}

The model introduced in this paper is a promising tool for separating overlapping components in complex mixtures of audio signals in the TF domain. Incorporating phase constraints within a complex NMF framework leads to satisfactory sounding results. Those constraints are based on signal modeling, which appears to be a suitable approach for exploiting the phase coherence properties that intrinsically lie within audio signals because of their physical nature. Promising results have been obtained for the source separation task, and significantly better results than with the traditional unconstrained complex NMF have been reached.

Further experiments should though be conducted on realistic music pieces, in order to assess the potential of the method for practical applications. In addition, further research could focus on the formulation of the problem itself. Indeed, incorporating the phase constraints as penalty terms in the cost function leads to an increase of the amount of local minima, and requires the tuning of the hyper parameters $\sigma_r$ and $\sigma_u$. Finally, such constraints could be refined, and completed with other properties, such as the modeling of onset phases or the use of time-invariant parameters~\cite{Kirchhoff2014}.

\newpage
\bibliographystyle{IEEEbib}
\bibliography{references_icassp2016}

\begin{thebibliography}{10}

\bibitem{Lee1999}
Daniel~D. Lee and H.~Sebastian Seung,
\newblock ``{Learning the parts of objects by non-negative matrix
  factorization},''
\newblock {\em Nature}, vol. 401, no. 6755, pp. 788--791, 1999.

\bibitem{Smaragdis2003}
Paris Smaragdis and Judith~C. Brown,
\newblock ``{Non-negative matrix factorization for polyphonic music
  transcription},''
\newblock in {\em {Proc. IEEE Workshop on Applications of Signal Processing to
  Audio and Acoustics (WASPAA)}}, New Paltz, NY, USA, October 2003.

\bibitem{Gerkmann2015}
Timo Gerkmann, Martin Krawczyk-Becker, and Jonathan {Le Roux},
\newblock ``{Phase Processing for Single-Channel Speech Enhancement: History
  and recent advances},''
\newblock {\em IEEE Signal Processing Magazine}, vol. 32, no. 2, pp. 55--66,
  March 2015.

\bibitem{Fevotte2009}
C{\'e}dric F{\'e}votte, Nancy Bertin, and Jean-Louis Durrieu,
\newblock ``{Nonnegative matrix factorization with the {Itakura-Saito}
  divergence: With application to music analysis},''
\newblock {\em Neural computation}, vol. 21, no. 3, pp. 793--830, March 2009.

\bibitem{Griffin1984}
Daniel Griffin and Jae~S. Lim,
\newblock ``{Signal estimation from modified short-time {F}ourier transform},''
\newblock {\em IEEE Transactions on Acoustics, Speech and Signal Processing},
  vol. 32, no. 2, pp. 236--243, April 1984.

\bibitem{Magron2015}
Paul Magron, Roland Badeau, and Bertrand David,
\newblock ``{Phase Reconstruction in {NMF} for audio source separation: An
  Insightful Benchmark},''
\newblock in {\em {Proc. IEEE International Conference on Acoustics, Speech and
  Signal Processing (ICASSP)}}, Brisbane, Australia, April 2015.

\bibitem{Badeau2014}
Roland Badeau and Mark~D. Plumbley,
\newblock ``{Multichannel High resolution {NMF} for modelling Convolutive
  Mixtures of Non-Stationary signals in the time-frequency domain},''
\newblock {\em IEEE Transactions on Audio Speech and Language Processing}, vol.
  22, no. 11, pp. 1670--1680, November 2014.

\bibitem{Kameoka2009}
Hirokazu Kameoka, Nobutaka Ono, Kunio Kashino, and Shigeki Sagayama,
\newblock ``{Complex {NMF}: A new sparse representation for acoustic
  signals},''
\newblock in {\em {Proc. IEEE International Conference on Acoustics, Speech and
  Signal Processing (ICASSP)}}, Taipei, Taiwan, April 2009.

\bibitem{McAuley1986}
R.~J. McAuley and T.~F. Quatieri,
\newblock ``{Speech analysis/Synthesis based on a sinusoidal representation},''
\newblock {\em IEEE Transactions on Acoustics, Speech and Signal Processing},
  vol. 34, no. 4, pp. 744--754, August 1986.

\bibitem{Flanagan1966}
J.~L. Flanagan and R.~M. Golden,
\newblock ``{Phase Vocoder},''
\newblock {\em Bell System Technical Journal}, vol. 45, pp. 1493--1509,
  November 1966.

\bibitem{Bronson2014}
James Bronson and Philippe Depalle,
\newblock ``{Phase constrained complex {NMF}: {Separating} overlapping partials
  in mixtures of harmonic musical sources},''
\newblock in {\em {Proc. IEEE International Conference on Acoustics, Speech and
  Signal Processing (ICASSP)}}, Florence, Italy, May 2014.

\bibitem{Krawczyk2014}
Martin Krawczyk and Timo Gerkmann,
\newblock ``{{STFT} Phase Reconstruction in Voiced Speech for an Improved
  Single-Channel Speech Enhancement},''
\newblock {\em IEEE/ACM Transactions on Audio, Speech, and Language
  Processing}, vol. 22, no. 12, pp. 1931--1940, December 2014.

\bibitem{Magron2015a}
Paul Magron, Roland Badeau, and Bertrand David,
\newblock ``{Phase reconstruction of spectrograms with linear unwrapping:
  application to audio signal restoration},''
\newblock in {\em {Proc. European Signal Processing Conference (EUSIPCO)}},
  Nice, France, August 2015.

\bibitem{Magron2015c}
Paul Magron, Roland Badeau, and Bertrand David,
\newblock ``{Phase reconstruction of spectrograms based on a model of repeated
  audio events},''
\newblock in {\em {Proc. IEEE Workshop on Applications of Signal Processing to
  Audio and Acoustics (WASPAA)}}, New Paltz, NY, USA, October 2015.

\bibitem{Hoyer2004}
Patrik~O. Hoyer,
\newblock ``{Non-negative {Matrix} {Factorization} with {Sparseness}
  {Constraints}},''
\newblock {\em Journal of Machine Learning Research}, vol. 5, pp. 1457--1469,
  2004.

\bibitem{Sawada2011}
Hiroshi Sawada, Hirokazu Kameoka, Shoko Araki, and Naonori Ueda,
\newblock ``{Formulations and algorithms for multichannel complex {NMF}},''
\newblock in {\em {Proc. IEEE International Conference on Acoustics, Speech and
  Signal Processing (ICASSP)}}, Prague, Czech Republic, May 2011, pp. 229--232.

\bibitem{LeRoux2009a}
Jonathan {Le Roux}, Hirokazu Kameoka, Emmanuel Vincent, Nobutaka Ono, Kunio
  Kashino, and Shigeki Sagayama,
\newblock ``{Complex {NMF} under spectrogram consistency constraints},''
\newblock in {\em {Proc. Acoustical Society of Japan Autumn Meeting}},
  Hukushima, Japan, September 2009.

\bibitem{Kirchhoff2014}
Holger Kirchhoff, Roland Badeau, and Simon Dixon,
\newblock ``{Towards complex matrix decomposition of spectrogram based on the
  relative phase offsets of harmonic sounds},''
\newblock in {\em {Proc. IEEE International Conference on Acoustics, Speech and
  Signal Processing (ICASSP)}}, Florence, Italy, May 2014.

\bibitem{Paulus2005}
Jouni Paulus and Virtanen Tuomas,
\newblock ``{Drum transcription with non-negative spectrogram factorisation},''
\newblock in {\em {Proc. European Signal Processing Conference (EUSIPCO)}},
  Antalya, Turkey, September 2005, IEEE, pp. 1--4.

\bibitem{Hua2004}
Yingbo Hua, Alex~B. Gershman, and Qi~Cheng,
\newblock {\em {High-resolution and robust signal processing}},
\newblock {Signal processing and communications}. Marcel Dekker, New York,
  2004.

\bibitem{Emiya2010a}
Valentin Emiya, Nancy Bertin, Bertrand David, and Roland Badeau,
\newblock ``{{MAPS} - {A} piano database for multipitch estimation and
  automatic transcription of music},''
\newblock Tech. {R}ep. 2010D017, T{\'e}l{\'e}com ParisTech, Paris, France, July
  2010.

\bibitem{Vincent2006}
Emmanuel. Vincent, R{\'e}mi Gribonval, and C{\'e}dric F{\'e}votte,
\newblock ``{Performance Measurement in Blind Audio Source Separation},''
\newblock {\em IEEE Transactions on Speech and Audio Processing}, vol. 14, no.
  4, pp. 1462--1469, July 2006.

\bibitem{Magron}
Paul Magron,
\newblock ``{Webpage},'' 2015,
\newblock "\url{http://perso.telecom-paristech.fr/~magron/}".

\end{thebibliography}

\end{document}